# ON THE POSSIBILITIES FOR SUPERLUMINAL SIGNALLING

## Alex Malikotsinas[1]

### Abstract

This paper shall define and discuss two types of quantum process - Disentangling and Entangling. The first type will be shown to contradict Unitarity and is therefore ruled out as a possible signalling process within standard linear Quantum Mechanics. The paper will argue that the second type –the Entangling process– is both allowed by the principles of QM, and can transmit superluminal signals. Tunnelling, it is argued, is an example of the Entangling process. A proposal will be made for addressing the objection to superluminal signalling by the Special Theory of Relativity.

### 1. The Disentangler

The Disentangler is a hypothetical linear device that acts at one end of an entangled pair of particles and brings about the following evolution:

$$|\Psi_{ent}\rangle = \frac{1}{\sqrt{2}}\left(|\Psi_1\rangle + |\Psi_2\rangle\right) \rightarrow |\Psi_{dis}\rangle = \alpha|\Psi_1\rangle + \beta|\Psi_2\rangle,$$

where $|\Psi_1\rangle$ and $|\Psi_2\rangle$ constitute an orthonormal basis set for the relevant Hilbert space of the entanglement, and by Unitarity $|\alpha|^2 + |\beta|^2 = 1$. We could view Unitarity as a quantum conservation principle. The conserved 'quantity' can be said to be *probability* or *total amplitude* or *normalisation*. A unitary operation transforms a normalised state-vector to another normalised state-vector [1].

Crucially we also require of the Disentangler that $|\alpha| \neq \frac{1}{\sqrt{2}}$ and $|\beta| \neq \frac{1}{\sqrt{2}}$. I will call this the *Signalling Assumption*.

This device shall be referred to as a Disentangler because it introduces a degree of disentanglement depending on the relative values of the coefficients α and β. Maximum disentanglement occurs when either α=0 or β=0. In this case the original entangled state is reduced - collapses completely - to a product state $|\Psi_1\rangle$ or $|\Psi_2\rangle$.

The action of the Disentangler is clearly seen with a photon-pair example:

$$|\Psi_{ent}\rangle = \frac{1}{\sqrt{2}}\left(|H_1 H_2\rangle + |V_1 V_2\rangle\right) \rightarrow |\Psi_{dis}\rangle = \alpha|H_1 H_2\rangle + \beta|V_1 V_2\rangle$$

where $|\alpha| \neq \frac{1}{\sqrt{2}}$ and $|\beta| \neq \frac{1}{\sqrt{2}}$, and $|\alpha|^2 + |\beta|^2 = 1$.

$H_i(V_i)$ denotes that the $i^{th}$ photon has horizontal (vertical) polarisation.

[1] alma@rmit.edu.au

If, for example, β=0, then $|\Psi_{dis}\rangle = \alpha|H_1H_2\rangle$. In this case the Disentangler causes the initial entangled state to always collapse to a product state comprising of two photons each possessing a definite horizontal polarisation. Similarly, for the case when α=0 the action of the Disentangler again assures a definite polarisation outcome since the final state is of the form $|\Psi_{dis}\rangle = \beta|V_1V_2\rangle$.

Such a device can send a signal. Given the initial state $|\Psi_{ent}\rangle = \frac{1}{\sqrt{2}}(|H_1H_2\rangle \pm |V_1V_2\rangle)$, the probability of obtaining the result $|H_2\rangle$ by performing a measurement on photon 2 (the Receiver's end) is exactly 0.5. However, following the action of the Disentangler at the Sender's end of the entanglement (photon 1), the state of the system becomes $|\Psi_{dis}\rangle$. The probability of obtaining a $|H_2\rangle$ for photon 2, at the Receiver's end, is now $|\alpha|^2 \neq 0.5$. That is, the Disentangler has altered the statistics one expects to collect by performing measurements on photon 2, which is at an arbitrary distance from the Disentangler device. In the limiting case when β=0, $|\alpha|^2 = 1$ and the signalling process reaches its maximum efficiency since the action of the device guarantees the result $|H_2\rangle$ at the Receiver's end.

The Disentangler therefore acts at one end of an entanglement, but causes the statistics collected at the other end to be altered and it can therefore send a signal. Such a signal will be superluminal provided the two events -the action of the Disentangler near photon 1 and the statistics collection near photon 2- are space-like separated.

Is such a device possible? It will be shown that the Disentangler is an impossible device within the context of standard *linear* QM because it violates Unitarity. The impossibility conclusion is in agreement with the large number of no-signalling theorems, which also argue that it is impossible to send signals by exploiting the properties of entangled pairs of particles [2,3,4]. But unlike the following proof, these other no-signalling theorems do not base their proof solely on a fundamental principle like Unitarity. Instead they rely on other assumptions like the commutativity of space-like separated operators [5,6].

## 2. Proof

Consider a device that brings about the following evolution by acting on one side of a correlated pair of photons:

$$|\Psi_{ent}\rangle \rightarrow |\Psi_{dis}\rangle$$

where $|\Psi_{ent}\rangle = \frac{1}{\sqrt{2}}(|H_1H_2\rangle + |V_1V_2\rangle)$,

$|\Psi_{dis}\rangle = \alpha|H_1H_2\rangle + \beta|V_1V_2\rangle$, and

$|\alpha| \neq \frac{1}{\sqrt{2}}, |\beta| \neq \frac{1}{\sqrt{2}}$ (Signalling Assumption).

Let the Device act on one side –photon 1. Assuming the device is linear, its action can be analysed in terms of its effect on each eigenstate:

$$|H_1 H_2\rangle \rightarrow (a|H_1\rangle + b|V_1\rangle)|H_2\rangle = a|H_1 H_2\rangle + b|V_1 H_2\rangle,$$

where by Unitarity $|a|^2 + |b|^2 = 1$, and

$$|V_1 V_2\rangle \rightarrow (c|H_1\rangle + d|V_1\rangle)|V_2\rangle = c|H_1 V_2\rangle + d|V_1 V_2\rangle,$$

where again by Unitarity $|c|^2 + |d|^2 = 1$.

By Linearity we have,

$$\frac{1}{\sqrt{2}}(|H_1 H_2\rangle + |V_1 V_2\rangle) \rightarrow \frac{1}{\sqrt{2}}(a|H_1 H_2\rangle + b|V_1 H_2\rangle + c|H_1 V_2\rangle + d|V_1 V_2\rangle).$$

Clearly, this final state will be of the required form $|\Psi_{dis}\rangle$, if $b = c = 0$ and $\frac{a}{\sqrt{2}} = \alpha, \frac{d}{\sqrt{2}} = \beta$.

Hence $a = \sqrt{2}\alpha$ and $d = \sqrt{2}\beta$.

Since by the Signalling Assumption $\alpha \neq \frac{1}{\sqrt{2}}$ and $\beta \neq \frac{1}{\sqrt{2}}$, it follows that $|a|^2 = 2|\alpha|^2 \neq 1$ and $|d|^2 = 2|\beta|^2 \neq 1$.

That is, $|a|^2 \neq 1$ and $|d|^2 \neq 1$.

However, since we require that $b = c = 0$, and by the initial assumption that the device obeys Unitarity, we obtain

$|a|^2 + |b|^2 = |a|^2 = 1$, and $|c|^2 + |d|^2 = |d|^2 = 1$, which results in a contradiction.

It may be concluded then, that the Disentangler is a device that transforms the state $|H_1 H_2\rangle$ (or $|V_1 V_2\rangle$) into the state $a|H_1 H_2\rangle$ (or $d|V_1 V_2\rangle$) where $|a|^2 \neq 1$ and $|d|^2 \neq 1$.

However, this contradicts Unitarity and hence the Disentangler must be an impossible device. Thus, there can be no device within linear Quantum Mechanics that can signal by reducing the degree of entanglement, or even by causing the complete collapse of an entanglement.

## 3. The Entangler

The Entangler is a device that transforms a non-entangled (product) state into an entangled one. Its action can be represented as follows:

$$|\Psi_1\rangle \to \frac{1}{\sqrt{2}}(|\Psi_1\rangle \pm |\Psi_2\rangle), \text{ and}$$

$$|\Psi_2\rangle \to \frac{i}{\sqrt{2}}(|\Psi_1\rangle \pm |\Psi_2\rangle)$$

where, $|\Psi_1\rangle$ and $|\Psi_2\rangle$ constitute an orthonormal basis set.

More generally, the action of the Entangler is: $|\Psi_1\rangle \to (a|\Psi_1\rangle + b|\Psi_2\rangle)$, where $|a|^2 + |b|^2 = 1$. The above case where $a = b = \frac{1}{\sqrt{2}}$ is a special one that makes the argument easier to follow.

The Entangler does not run afoul of either Unitarity or Linearity and is therefore possible in principle.

Proof:
$$|\Psi_1\rangle \to \frac{1}{\sqrt{2}}(|\Psi_1\rangle \pm |\Psi_2\rangle),$$
$$|\Psi_2\rangle \to \frac{i}{\sqrt{2}}(|\Psi_1\rangle \pm |\Psi_2\rangle).$$

By Linearity; $\frac{1}{\sqrt{2}}(|\Psi_1\rangle + |\Psi_2\rangle) \to \frac{1+i}{2}(|\Psi_1\rangle \pm |\Psi_2\rangle),$

and $2\left|\frac{1+i}{2}\right|^2 = 1$ which satisfies Unitarity.

Furthermore, such a device will signal. The signal will be potentially superluminal provided the device brings about the above evolution instantly and by acting on one side only.

For example, consider the initial two-photon product state $|H_1 H_2\rangle$.

The Entangler transforms this state into $\frac{1}{\sqrt{2}}(|H_1 H_2\rangle + |V_1 V_2\rangle)$ by acting on one of the photons (photon 1). The action of the device then causes the probability of obtaining the result $|H_2\rangle$ on the distant photon 2 to change from 1 to ½ instantly. This change in probability can be translated into a signal that is sent from the region of action of the device (photon 1), to the region where photon 2 is detected. Such a signal will be superluminal if the separation of the photons can be made arbitrarily large, and the total time it takes the device to affect the change of state is short.

However, consider the physical processes involved in the action of the Entangler: Take two uncorrelated photons, each in a known polarisation state, and separate them by an arbitrary distance. The Entangler then acts on *only one* of the two photons, but its action causes the two photons to become entangled thereby altering the expected statistics of polarisation measurements on the distant photon. Such a device can clearly signal, even superluminally, but one could object that its action appears miraculous. The Entangler, by acting on a local photon affects a far-distant photon via some seemingly mysterious instantaneous action-at-a-distance so as to entangle it with the first photon. Despite its seemingly miraculous nature, the Entangler is a physically possible device. The following section argues that we already have an example of such entangling-at-a-distance in the phenomenon of 'tunnelling'- a complete misnomer since the particle is not deemed to be travelling (tunnelling) through the barrier.

### 4. Tunnelling as Entanglement-at-a-distance

It will be argued that tunnelling is completely analogous to Entangling as it was defined in the previous section, and therefore just one instance of a more general Entangling process.

Consider two boxes labelled $X_1$ and $X_2$. If a particle is initially confined inside box $X_1$ its state can be written as:

$$|\Psi_1\rangle = |X_1 = 1\rangle |X_2 = 0\rangle = |X_1 = 1, X_2 = 0\rangle,$$

where $|X_i = 1, 0\rangle$ denotes that the $i^{th}$ box is either occupied (1) or empty (0).

$|\Psi_2\rangle = |X_1 = 0, X_2 = 1\rangle$ is the other basis state, which signifies that the particle is in box $X_2$ and box $X_1$ is empty.

Now let the initial state be $|\Psi_1\rangle$. The particle is deemed to be in $X_1$, and the probability of finding it in $X_2$ is zero. Then if a local action in the vicinity of $X_1$ allows the particle to 'tunnel' through to $X_2$, the final state will be an entangled one of the form:

$$|\Psi_{ent}\rangle = \frac{1}{\sqrt{2}}(|\Psi_1\rangle \pm |\Psi_2\rangle) = \frac{1}{\sqrt{2}}(|X_1 = 1, X_2 = 0\rangle \pm |X_1 = 0, X_2 = 1\rangle)$$

(Assuming no detection measurements are made on either box at this stage).

This $|\Psi_{ent}\rangle$ state has exactly the same form as the entangled photon-polarisation state discussed above. $|\Psi_{ent}\rangle$ in this case represents the entanglement of position states. The fact that we are dealing with a single particle should not detract from the fact that the form of $|\Psi_{ent}\rangle$ represents an entanglement of states – in this case position states.

In this entangled state, the probability of finding the particle in $X_2$ is 0.5. Therefore a signal may be sent from the first box to the second because the local action on the first box alters the statistics that one can collect at the second box. For example, the two boxes can be connected by a tube with a diameter smaller than the wavelength of the particle. This is the entangling ('tunnelling') medium. Undersized waveguides allow the propagation of evanescent modes and hence tunnelling. Nimtz makes this point in his paper where he presents the results of his superluminal tunnelling experiments [7]. The tube is initially blocked at the sender's end – $X_1$ - by an 'infinite' potential which can be removed to allow tunnelling. The removal of the blockage in the entangling medium allows tunnelling to take place and the establishment of the entangled state shown above. Another possibility is the injection of the particle into $X_1$ while the two boxes are permanently connected via an open tunnelling medium, such as the undersized waveguide in the previous example. In this case, also, the pre-tunnelling probability of detecting the particle in $X_2$ is zero. Immediately after the injection and tunnelling this probability is non-zero, and this can be translated into a signal from $X_1$ to $X_2$. Obviously, such a device cannot send a reliable signal on a single particle basis. One needs many particles prepared in the same state, or a parallel array of similar devices so that the statistics collected at the receiver's end give an unambiguous result before and after the sender's action. The Entangler is therefore a physically possible device at least in the case of the entanglement of position states of a single particle.

It is important to note that the conversion of the altered expectation values to a signal involves the process of measurement at the receiver's end, and therefore the collapse of the entanglement. The 'collapse of the wave packet' and 'measurement' remain unresolved issues for standard linear quantum mechanics. Apart from some brief comments later on, this paper will not discuss these issues. The important point here is that the action of the Entangler that brings about the change in expectation values and therefore makes signalling possible, is completely described by linear quantum mechanics.

A recent letter to *Nature* by Stenner *et.al.* shows that despite the superluminality of the group velocity of a pulse traversing an anomalous dispersion medium, the speed of 'information'-defined by the authors as a discontinuity in the amplitude of the pulse - is always lower than the vacuum speed of light [8]. These results could be viewed as contradicting the argument in this paper that superluminal signals are possible. Nimtz, whose tunnelling experiments have shown superluminality, disputes the interpretation of the experimental results reported in the *Nature* article and concludes that this experiment by Stenner *et.al.* does not prove the impossibility of superluminal information transfer [9]. A detailed study of this dispute is not necessary at this juncture, but is important to stress that my argument above is concerned solely with tunnelling-type processes, not anomalous dispersion. This distinction is not adequately appreciated in the literature and this confounds the issues involved.

## 5. Other Possibilities

By exact analogy to particle tunnelling we could have two spin-½ particles both in a definite spin state and confined to two distant regions. The two regions must be connected by a spin-tunnelling medium, which is blocked locally at the sender's end.

The removal of the blocking potential by the sender allows tunnelling which transforms an initial state of the form $|\Psi_1\rangle = |+\rangle_1 |-\rangle_2$ (or $|\Psi_2\rangle = |-\rangle_1 |+\rangle_2$), into:

$$|\Psi_{ent}\rangle = \frac{1}{\sqrt{2}}\left(|+\rangle_1|-\rangle_2 \pm |-\rangle_1|+\rangle_2\right),$$

where $|\pm\rangle_i$ indicates that the i[th] particle has spin ±½.

Again a signal can be sent because the sender's action on particle 1 alters the probability of obtaining a +1/2 (or -1/2) at the receiver's end (particle 2). Before tunnelling, the receiver's particle is in a definite (and known) spin state given by either $|\Psi_1\rangle$ or $|\Psi_2\rangle$ above. If the initial state is $|\Psi_1\rangle$ then the probability of obtaining a -1/2 for particle 2 at the receiver's end is 1. Following the activation of the entangling medium and the formation of the entangled state the probability of obtaining a -1/2 becomes 0.5. This change in probability at the receiver's end can be transformed into a signal. If the two regions that are being entangled are space-like separated, then the signal that this process sends across the two regions will be superluminal.

The spin and particle tunnelling examples are exactly analogous and they constitute instances of the operation of the Entangler device. They point to the possibility of a photon-polarisation Entangler and they provide hints as to the key features of such a device. A photon-polarisation device will have to rely on some medium (or potential), which allows the polarisation state of a photon to tunnel through. The two photons whose polarisation is to be entangled must be connected by the appropriate medium/potential. In general then, the Entangler is a tunnelling-type device that relies on an appropriate medium/potential in order to cause the entanglement of states in two distant regions. Of special interest is the case where a particle that is initially confined inside a box is suddenly freed by removing the infinite potential walls surrounding it. Quantum Mechanics predicts that instantly the probability of the particle being detected anywhere outside the box is raised from zero to a finite value [10,11]. Superluminal signalling again becomes possible in principle. This is an instance where free space provides the medium for the Entangler.

Are there such entangling media/potentials that can also be extended over arbitrary distances? The tunnelling of particles is an established fact but so far it is shown to occur over relatively short distances and the superluminality of the phenomenon is still a hotly debated topic [12]. However, the analysis above shows that the tunnelling of particles is just a special case of what I call the Entangling process. There is no principle of Quantum Mechanics that either forbids such a process or restricts its range of application.

### 6. The Relativity Objection

The only objection to the possibility of such a device comes from the Special Theory of Relativity(STR). The objection is that, according to STR if superluminal signals are allowed then causality will be violated since the effect will precede the cause in some frames of reference [8,13]. This is both a well known and accepted argument which would definitely rule out the possibility of superluminal signals provided Relativity theory is taken as the final word on the structure of space-time. The argument may be summarised as follows:

The conjunction of: (i) Quantum Mechanics (standard non-relativistic), and (ii) Superluminal Signals, and (iii) the Special Theory of Relativity, and (iv) Causality, leads to intolerable causal paradoxes.

The standard way of avoiding the causal paradoxes has so far consisted in ruling out superluminal signals. As I have argued elsewhere this is not the only possible response [14]. I would raise it as a possibility at this stage that the weakest premise in the above argument is STR. I claim that it is possible to have standard Quantum Mechanics with Superluminal Signalling without violating Causality, provided we modify our views of space-time. In the absence of a generally accepted instance of superluminal signalling, there is no knockdown argument for a revision of our views on space-time. Nevertheless, there is a lot of circumstantial evidence that points to the need for such a revision. One instance of such evidence is provided by the Collapse of the Wave Packet. This collapse is generally taken to be instantaneous but has resisted all attempts to 'relativise' it, that is, find a covariant expression for it. There is no suggestion here that the instantaneity of the collapse can by itself lead to superluminal signalling. My own argument above against the Disentangler device shows the impossibility of such signalling. The point is that STR cannot accommodate this feature of Quantum Mechanics. Aharonov and Albert in a couple of papers 20 years ago showed the serious inadequacies of all previous attempts to deal with the problem [15,16]. They even propose their own solution, which also fails to reconcile Relativity with this aspect of Quantum Mechanics. This is not the place to discuss the intricacies of their proposal, but suffice it to say that they come close to destroying the notion of statehood in Quantum Mechanics. The state of a quantum system is no longer a function of space-time and depends in a non-trivial way upon the state of motion of the observer. Two inertial observers with a relative velocity will ascribe states $\psi_1$ and $\psi_2$ respectively to the *same* system at the *same* space-time point but $\psi_1$ and $\psi_2$ are *not* Lorentz transforms of each other [17]. It is ironic that a proposal designed to save Relativity actually undermines it by claiming that the state vector is not a covariant function of space-time.

Given the serious difficulties in producing a covariant explanation of quantum superluminal phenomena like the instantaneous collapse, a more promising approach might be to focus on possible extensions of Relativity. Such extension should aim at saving standard Quantum Mechanics, the superluminal phenomena it predicts (including signalling), as well as causality.

## 7. Conclusion

The Disentangler (a hypothetical superluminal signalling device) is impossible within linear Quantum Mechanics because it contradicts Unitarity. There are non-linear versions of Quantum Mechanics that could render the Disentangler possible [18]; however this issue is beyond the scope of this paper. The *Entangler* on the other hand does not contradict any principles of Quantum Mechanics. The well understood and accepted phenomenon of tunnelling is an instance of the operation of the Entangler and this shows that it is a physically possible device. The Entangler can send a signal that can be superluminal. The conjunction of superluminal signals and the Special

Theory of Relativity violates causality. One way of saving superluminal signalling and other quantum superluminal phenomena without violating causality is by extending our relativistic notions of space-time. Such a modified space-time structure is proposed in a follow-up paper based on [14].

**Acknowledgments:**

I am grateful to Martin Leckey for his very useful and insightful comments on previous drafts of this paper. Many thanks also to Neil Thomason and Keith Hutchison for numerous discussions on issues relating to this paper.